# A New Approach to Linear Estimation Problem in Multi-user Massive MIMO Systems

**Muhammad Ali Raza Anjum**
Army Public College of Management and Sciences, Rawalpindi, Pakistan
E-mail: ali.raza.anjum@gmail.com

***Abstract***
*A novel approach for solving linear estimation problem in multi-user massive MIMO systems is proposed. In this approach, the difficulty of matrix inversion is attributed to the incomplete definition of the dot product. The general definition of dot product implies that the columns of channel matrix are always orthogonal whereas, in practice, they may be not. If the latter information can be incorporated into dot product, then the unknowns can be directly computed from projections without inverting the channel matrix. By doing so, the proposed method is able to achieve an exact solution with a 25% reduction in computational complexity as compared to the QR method. Proposed method is stable, offers an extra flexibility of computing any single unknown, and can be implemented in just twelve lines of code.*

## 1. Introduction

Multiple Input Multiple Output (MIMO) systems incorporate multiple antennas at the transmitter and the receiver to improve the data rate of wireless communication systems [1, 2]. However, the ever growing desire for the increased data rates can hardly be satiated and there is a demand to increase the data rates even further. As a result, the researchers are aiming to achieve the asymptotic limits in capacity. Recently, it has been demonstrated that very high capacities are achievable on forward and reverse links as the number of transmit antennas approaches infinity [3]. Such systems in which a relatively large number of base station antennas serve a large number of users are known as Massive MIMO systems (M-MIMO) [4].

M-MIMO systems provide various advantages over the traditional MIMO systems. These systems promise higher data rates, higher antenna resolution, lower error probabilities, lower thermal noise, and lower transmit power per antenna. These advantages can be attributed to an overall averaging behavior of these systems. But these advantages become even more impressive in the multi-user scenario where the base station transmits to several users simultaneously. However, multi-user M-MIMO systems incur costs of their own: including increase in hardware, increase in power consumption, and increase in physical spacing. Eventually, the transceiver becomes complex as well [3-6].

Transceiver complexity is currently an issue of great concern in M-MIMO systems. It has been shown that for the point to point scenario, the decoding complexity of the receiver alone grows exponentially with increase in the number of transmit antennas [7]. In case of a multi-user scenario, the transmitter becomes complex as well because the transmitter now requires advanced coding schemes to manage simultaneous transmission of information to multiple users. Former operation, known as decoding, and the latter one, called precoding, generally comprise the transceiver operation [8].

For a large number of transmit antennas, linear decoders and precoders have proven almost optimal [9-11]. These linear precoders/decoders require the inversion of a potentially large channel matrix. For example, a practical precoding method is the Zero-forcing (ZF) precoding method which computes the precoding matrix by forming the pseudoinverse of the channel matrix [9]. Similar to linear precoding, a simple linear decoder is the MMSE decoder which computes the decoding matrix by again forming the pseudoinverse of the channel matrix [12]. If the dimensions of a system are very large, inverting the channel matrix becomes a cumbersome task [4].

Therefore, there is a need to eliminate the outright inversion. Various methods for approximate matrix inversion have been proposed in literature: including Cayley-Hamilton theorem method, Neumann series expansion method, QR decomposition method, random matrix methods, and methods based on polynomial and truncated polynomial filters [13-21]. These methods still require a lot of computational effort and some of them have proven to be even more complex than the direct inversion. However, it is important to note that the challenge is fundamentally different in at least three possible ways. To begin with, the channel matrix does not have a structure. Secondly, instead of being deterministic, it is random. Finally, there is no guarantee of sparsity. So we can dispense with the traditional methods for solving the linear estimation problem [22].

With this in view, we propose a novel method to solve such large linear systems. The proposed method treats the linear estimation problem as a coordinate transformation problem. By doing so, the method is able to achieve zero norms in error and residue in contrast to the state of the art iterative methods for large systems – LSQR, LSMR, and Kaczmarz that achieve much higher norms - and a 25% reduction in computation complexity when compared to the QR method, the leading protagonist in exact methods [23-25]. Also by employing the proposed method, a particular unknown, which may be the only information required by a single user, can be computed without evaluating the entire solution vector while no such flexibility is available in the traditional methods. Furthermore, the method is stable and light on computation as well as it only relies on multiplication with simple rank-one projection matrices for obtaining the solution. Finally, the method can be programmed in just ten lines of codes.

Rest of the article is organized as follows. Section 2 begins with a system model and defines the underlying problem. Basic idea behind the proposed method is explored in Section 3. The method is then explained in Section 4. Section 5 contains its complete proof. Section 6 provides the proof of the choice of reflection matrices. A complete algorithm for step by step solution of the linear estimation problem according to the proposed method is outlined in Section 7. This section also includes a user-friendly code for the algorithm. Comparison of the proposed method with QR decomposition is carried out in section 8. The process of generation of inverse vectors is also discussed in this section. Computational complexity of the proposed algorithm is analyzed in Section 9 which is followed by its stability analysis in Section 10. Section 11 concludes the article in which the comparison of the proposed method is carried out with the state of the art methods available in literature for solving large systems.

## 2. Problem

A narrowband memoryless MIMO channel with $n$ transmit and $m$ receive antennas can be modeled as a system of linear equations [13].

$$A_n x = b \tag{1}$$

With,

$$A_n = \begin{bmatrix} a^T{}_1 \\ a^T{}_2 \\ \vdots \\ a^T{}_n \end{bmatrix}^T \qquad x = \begin{bmatrix} x_1 \\ x_2 \\ \vdots \\ x_n \end{bmatrix} \qquad b = \begin{bmatrix} b_1 \\ b_2 \\ \vdots \\ b_m \end{bmatrix} \tag{2}$$

$A_n$ is a large, random and non-sparse matrix. $x$ is the input vector. $b$ is the output vector. Zero Forcing (ZF) estimate $\hat{x}$ of the input vector is:

$$\hat{x} = \left(A_n{}^T A_n\right)^{-1} A_n{}^T b \tag{3}$$

Recovery of $x$ requires the inversion of a potentially large matrix $A_n{}^T A_n$.

## 3. Basic Idea

Linear estimation problem presented by Equation (3) can be viewed as a coordinate transformation problem. $b$ can be envisaged as the vector that undergoes the coordinate transformation, $x$ represents its coordinates in the new coordinate system, and $A_n$ contains the basis vectors for the new coordinate system [26]. The basis vectors can be orthogonal or non-orthogonal. For the first case, we are lead to the classic Least Squares (LS) solution depicted in Figure 1.

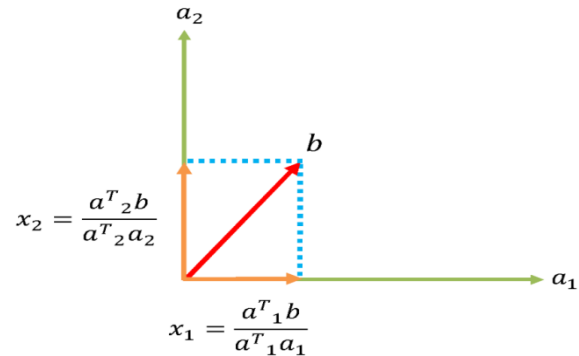

Figure 1. Computation of unknowns in an orthogonal coordinate system

In this case, the estimates $x_1, x_2, \ldots, x_n$ are the projections of $b$ on the respective basis vectors.

$$\begin{bmatrix} x_1 = \dfrac{a^T_1 b}{a^T_1 a_1} \\ x_2 = \dfrac{a^T_2 b}{a^T_2 a_2} \\ \vdots \\ x_n = \dfrac{a^T_n b}{a^T_n a_n} \end{bmatrix} \tag{4}$$

When the basis vectors are not orthogonal, $x$ is computed by forming the inverse of $A_n$ matrix. This is precisely what we desire to avoid. We would like to compute $x$ from the direct estimate of the projections. In order to do so, we focus our attention on Figure 2 for the case of non-orthogonal basis vectors. As can be observed from figure, employing the definition of dot product in Equation (4) will yield much longer estimates than the actual ones. Reason for these incorrect estimates can be attributed to the presence of identity matrix in the general definition of the dot product. We demonstrate this fact by re-writing the first term in Equation (4).

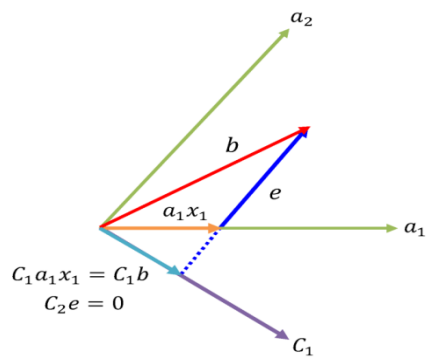

Figure 2. Computation of first unknown in a non-orthogonal coordinate system

$$x_1 = \frac{a^T_1 I b}{a^T_1 I a_1} \tag{5}$$

Identity matrix $I$ depicts a set of orthogonal basis vectors. As they are not, we can replace the $I$ matrix in Equation (5) by another matrix, say $C_1$.

$$x_1 = \frac{a^T_1 C_1 b}{a^T_1 C_1 a_1} \tag{6}$$

To solve for $x_1$, we re-arrange Equation (6) a bit.

$$a^T_1 C_1 a_1 x_1 = a^T_1 C_1 b \tag{7}$$

Or,

$$C_1 a_1 x_1 = C_1 b \tag{8}$$

Equation (8) requires the vectors $a_1 x_1$ and $b$ to have equal projections on a plane spanned by the column space of $C_1$. We observe from Figure 2 that $a_1 x_1$ and $b$ have equal projections on a plane orthogonal to the second basis vector $a_2$. $a_1 x_1$ and $b$ can be projected on such plane by a projection matrix of the form:

$$C_1 = \left[ I - \frac{a_2 a_2^T}{a_2^T a_2} \right] \tag{9}$$

We can substitute $C_1$ in Equation (5) to solve for $x_1$.

$$x_1 = \frac{a^T_1 \left[ I - \frac{a_2 a_2^T}{a_2^T a_2} \right] b}{a^T_1 \left[ I - \frac{a_2 a_2^T}{a_2^T a_2} \right] a_1} \tag{10}$$

Similarly for $x_2$,

$$x_2 = \frac{a^T_2 \left[ I - \frac{a_1 a_1^T}{a_1^T a_1} \right] b}{a^T_2 \left[ I - \frac{a_1 a_1^T}{a_1^T a_1} \right] a_2} = \frac{a^T_2 C_2 b}{a^T_2 C_2 a_2} \tag{11}$$

$C_2$ will project $a_2 x_2$ and $b$ on a plane orthogonal to $a_1$. Using Equation (10) and (11), we can directly solve for $x_1$ without inverting the matrix $A_n$. Equation (10) and (11) also have a connotation that $x_1$ can be solved independently of $x_2$.

### 4. Method

Now we present the method for solving the general problem of the form $A_n x = b$. We begin by expanding Equation (3),

$$[a_1 \quad a_2 \quad \ldots \quad a_{n-1} \quad a_n] x = b \tag{12}$$

Multiplying Equation (12) by a matrix $R_n$,

$$[R_n a_1 \quad R_n a_2 \quad \ldots \quad R_n a_{n-1} \quad R_n a_n] x = R_n b \tag{13}$$

In order for $R_n a_n$ to be zero,

$$R_n = \left[ I - \frac{a_n a_n^T}{a_n^T a_n} \right] \tag{14}$$

Equation (13) becomes,

$$[R_n a_1 \quad R_n a_2 \quad ... \quad R_n a_{n-1} \quad 0] x = R_n b \tag{15}$$

Multiplying Equation (15) by a matrix $R_{n-1}$,

$$[R_{n-1} R_n a_1 \quad R_{n-1} R_n a_2 \quad ... \quad R_{n-1} R_n a_{n-1} \quad 0] x = R_{n-1} R_n b \tag{16}$$

In order for $R_n a_{n-1}$ to be zero,

$$R_{n-1} = \left[ I - \frac{(R_n a_{n-1})(R_n a_{n-1})^T}{(R_n a_{n-1})^T (R_n a_{n-1})} \right] \tag{17}$$

Equation (16) becomes,

$$[R_{n-1} R_n a_1 \quad R_{n-1} R_n a_2 \quad ... \quad 0 \quad 0] x = R_{n-1} R_n b \tag{18}$$

Continuing in the same fashion,

$$[R_2 ... R_{n-1} R_n a_1 \quad 0 \quad ... \quad 0 \quad 0] x = R_2 ... R_{n-1} R_n b \tag{91}$$

or,

$$R_2 ... R_{n-1} R_n a_1 x_1 = R_2 ... R_{n-1} R_n b \tag{20}$$

Multiplying both sides by $a_1^T$,

$$a_1^T R_2 ... R_{n-1} R_n a_1 x_1 = a_1^T R_2 ... R_{n-1} R_n b \tag{21}$$

Or,

$$x_1 = \frac{a_1^T C_1 b}{a_1^T C_1 a_1} \tag{22}$$

With,

$$C_1 = R_2 ... R_{n-1} R_n \tag{23}$$

For the $k$-th unknown,

$$x_k = \frac{a_k^T C_k b}{a_k^T C_k a_k} \tag{24}$$

With,

$$C_k = R_1 R_2 ... R_i ... R_{n-1} R_n \; \exists \; i \neq k \tag{25}$$

## 5. Proof

In this section, we provide the proof of the proposed method. We begin the proof by writing the basic least squares problem.

$$A^T_n A_n x = A^T_n b \tag{26}$$

Expanding Equation (26),

$$\begin{bmatrix} a^T_1 \\ a^T_2 \\ \vdots \\ a^T_{n-1} \\ a^T_n \end{bmatrix} [a_1 \quad a_2 \quad \ldots \quad a_{n-1} \quad a_n] x = \begin{bmatrix} a^T_1 \\ a^T_2 \\ \vdots \\ a^T_{n-1} \\ a^T_n \end{bmatrix} b \tag{27}$$

$$\begin{bmatrix} a^T_1 a_1 & a^T_1 a_2 & \ldots & a^T_1 a_{n-1} & a^T_1 a_n \\ a^T_2 a_1 & a^T_2 a_2 & \ldots & a^T_2 a_{n-1} & a^T_2 a_n \\ \vdots & \vdots & \ddots & \vdots & \vdots \\ a^T_{n-1} a_1 & a^T_{n-1} a_2 & \ldots & a^T_{n-1} a_{n-1} & a^T_{n-1} a_n \\ a^T_n a_1 & a^T_n a_2 & \ldots & a^T_n a_{n-1} & a^T_n a_n \end{bmatrix} x = \begin{bmatrix} a^T_1 b \\ a^T_2 b \\ \vdots \\ a^T_{n-1} b \\ a^T_n b \end{bmatrix} \tag{28}$$

If columns of $A_n$ are orthonormal, then Equation (28) becomes,

$$\begin{bmatrix} 1 & 0 & \ldots & 0 & 0 \\ 0 & 1 & \ldots & 0 & 0 \\ \vdots & \vdots & \ddots & \vdots & \vdots \\ 0 & 0 & \ldots & 1 & 0 \\ 0 & 0 & \ldots & 0 & 1 \end{bmatrix} x = \begin{bmatrix} a^T_1 b \\ a^T_2 b \\ \vdots \\ a^T_{n-1} b \\ a^T_n b \end{bmatrix} \tag{29}$$

$$x = \begin{bmatrix} a^T_1 b \\ a^T_2 b \\ \vdots \\ a^T_{n-1} b \\ a^T_n b \end{bmatrix} \tag{30}$$

If columns of $A_n$ are orthogonal, then Equation (28) becomes,

$$\begin{bmatrix} a^T_1 a_1 & 0 & \ldots & 0 & 0 \\ 0 & a^T_2 a_2 & \ldots & 0 & 0 \\ \vdots & \vdots & \ddots & \vdots & \vdots \\ 0 & 0 & \ldots & a^T_{n-1} a_{n-1} & 0 \\ 0 & 0 & \ldots & 0 & a^T_n a_n \end{bmatrix} x = \begin{bmatrix} a^T_1 b \\ a^T_2 b \\ \vdots \\ a^T_{n-1} b \\ a^T_n b \end{bmatrix} \tag{31}$$

$$x = \begin{bmatrix} \frac{a^T_1 b}{a^T_1 a_1} \\ \frac{a^T_2 b}{a^T_2 a_2} \\ \vdots \\ \frac{a^T_{n-1} b}{a^T_{n-1} a_{n-1}} \\ \frac{a^T_n b}{a^T_n a_n} \end{bmatrix} \tag{32}$$

If columns of $A_n$ are neither orthonormal nor orthogonal, then the off-diagonal terms in Equation (28) prevent the direct solution. If we can somehow eliminate these terms, then a direct solution of the form (32) is possible. In order to do that, we insert a block matrix $C$ matrix in Equation (26) and re-write it as:

$$A^T_n C A_n x = A^T_n C b \tag{33}$$

Expanding (33),

$$\begin{bmatrix} a^T_1 \\ a^T_2 \\ \vdots \\ a^T_{n-1} \\ a^T_n \end{bmatrix} \begin{bmatrix} C_1 \\ C_2 \\ \vdots \\ C_{n-1} \\ C_n \end{bmatrix} [a_1 \quad a_2 \quad \ldots \quad a_{n-1} \quad a_n] x = \begin{bmatrix} a^T_1 \\ a^T_2 \\ \vdots \\ a^T_{n-1} \\ a^T_n \end{bmatrix} \begin{bmatrix} C_1 \\ C_2 \\ \vdots \\ C_{n-1} \\ C_n \end{bmatrix} b \tag{34}$$

$$\begin{bmatrix} a^T_1 \\ a^T_2 \\ \vdots \\ a^T_{n-1} \\ a^T_n \end{bmatrix} \begin{bmatrix} C_1 a_1 & C_1 a_2 & \ldots & C_1 a_{n-1} & C_1 a_n \\ C_2 a_1 & C_2 a_2 & \ldots & C_2 a_{n-1} & C_2 a_n \\ \vdots & \vdots & \ddots & \vdots & \vdots \\ C_{n-1} a_1 & C_{n-1} a_2 & \ldots & C_{n-1} a_{n-1} & C_{n-1} a_n \\ C_n a_1 & C_n a_2 & \ldots & C_n a_{n-1} & C_n a_n \end{bmatrix} x = \begin{bmatrix} a^T_1 \\ a^T_2 \\ \vdots \\ a^T_{n-1} \\ a^T_n \end{bmatrix} \begin{bmatrix} C_1 b \\ C_2 b \\ \vdots \\ C_{n-1} b \\ C_n b \end{bmatrix} \tag{35}$$

If only,

$$C_i a_j \neq 0 \; \exists \; i \neq j \tag{36}$$

Then Equation (35) takes the form,

$$\begin{bmatrix} a^T_1 C_1 a_1 & 0 & \ldots & 0 & 0 \\ 0 & a^T_2 C_2 a_2 & \ldots & 0 & 0 \\ \vdots & \vdots & \ddots & \vdots & \vdots \\ 0 & 0 & \ldots & a^T_{n-1} C_{n-1} a_{n-1} & 0 \\ 0 & 0 & \ldots & 0 & a^T_n C_n a_n \end{bmatrix} x = \begin{bmatrix} a^T_1 C_1 b \\ a^T_2 C_2 b \\ \vdots \\ a^T_{n-1} C_{n-1} b \\ a^T_n C_n b \end{bmatrix} \tag{37}$$

And the solution becomes,

$$x = \begin{bmatrix} \frac{a^T_1 C_1 b}{a^T_1 C_1 a_1} \\ \frac{a^T_2 C_2 b}{a^T_2 C_2 a_2} \\ \vdots \\ \frac{a^T_{n-1} C_{n-1} b}{a^T_{n-1} C_{n-1} a_{n-1}} \\ \frac{a^T_n C_n b}{a^T_n C_n a_n} \end{bmatrix} \tag{38}$$

For Equation (38) to hold, we can choose $C_i$ to be:

$$C_i = R_1 R_2 \ldots R_j \ldots R_{n-1} R_n \exists \; j \neq i \tag{39}$$

$R$ is a projection matrix.

$$R = \left[ I - \frac{aa^T}{a^T a} \right] \tag{40}$$

$R$ projects a vector on an axis that is orthogonal to the plane defined by $a$. An important property of $R$ matrix is that the product $Ra$ is zero. This is because the projection of any vector on an axis orthogonal to itself is always zero. Hence, the product $Ra$ is always zero.

$$Ra = \left[ I - \frac{aa^T}{a^T a} \right] a = a - \frac{aa^T a}{a^T a} = a - a = 0 \tag{41}$$

This property of $R$ matrix can be can be used to eliminate the off-diagonal entries in Equation (1).

$$\begin{bmatrix} a_1 & a_2 & \ldots & a_{n-1} & a_n \\ a_1 & a_2 & \ldots & a_{n-1} & a_n \\ \vdots & \vdots & \ddots & \vdots & \vdots \\ a_1 & a_2 & \ldots & a_{n-1} & a_n \\ a_1 & a_2 & \ldots & a_{n-1} & a_n \end{bmatrix} x = \begin{bmatrix} b \\ b \\ \vdots \\ b \\ b \end{bmatrix} \tag{42}$$

Starting with the last term in first row on the left hand side of Eq. (42), we multiply the top row with $R_n$.

$$\begin{bmatrix} R_n a_1 & R_n a_2 & \dots & R_n a_{n-1} & R_n a_n \\ a_1 & a_2 & \dots & a_{n-1} & a_n \\ \vdots & \vdots & \ddots & \vdots & \vdots \\ a_1 & a_2 & \dots & a_{n-1} & a_n \\ a_1 & a_2 & \dots & a_{n-1} & a_n \end{bmatrix} x = \begin{bmatrix} R_n b \\ b \\ \vdots \\ b \\ b \end{bmatrix} \quad (43)$$

With,

$$R_n = \left[ I - \frac{a_n a^T_n}{a^T_n a_n} \right] \quad (44)$$

Equation (43) becomes,

$$\begin{bmatrix} R_n a_1 & R_n a_2 & \dots & R_n a_{n-1} & 0 \\ a_1 & a_2 & \dots & a_{n-1} & a_n \\ \vdots & \vdots & \ddots & \vdots & \vdots \\ a_1 & a_2 & \dots & a_{n-1} & a_n \\ a_1 & a_2 & \dots & a_{n-1} & a_n \end{bmatrix} x = \begin{bmatrix} R_n b \\ b \\ \vdots \\ b \\ b \end{bmatrix} \quad (45)$$

Similarly, to eliminate $R_n a_{n-1}$ we multiply the top row in Equation (45) with $R_{n-1}$,

$$\begin{bmatrix} R_{n-1} R_n a_1 & R_{n-1} R_n a_2 & \dots & 0 & 0 \\ a_1 & a_2 & \dots & a_{n-1} & a_n \\ \vdots & \vdots & \ddots & \vdots & \vdots \\ a_1 & a_2 & \dots & a_{n-1} & a_n \\ a_1 & a_2 & \dots & a_{n-1} & a_n \end{bmatrix} x = \begin{bmatrix} R_{n-1} R_n b \\ b \\ \vdots \\ b \\ b \end{bmatrix} \quad (46)$$

Such that,

$$R_{n-1} = \left[ I - \frac{(R_n a_{n-1})(R_n a_{n-1})^T}{(R_n a_{n-1})^T (R_n a_{n-1})} \right] \quad (47)$$

We continue this multiplication in this manner until all the off diagonal entries in the first row are removed.

$$\begin{bmatrix} C_1 a_1 & 0 & \dots & 0 & 0 \\ a_1 & a_2 & \dots & a_{n-1} & a_n \\ \vdots & \vdots & \ddots & \vdots & \vdots \\ a_1 & a_2 & \dots & a_{n-1} & a_n \\ a_1 & a_2 & \dots & a_{n-1} & a_n \end{bmatrix} x = \begin{bmatrix} C_1 b \\ b \\ \vdots \\ b \\ b \end{bmatrix} \quad (48)$$

With $C_1$,

$$C_1 = R_2 \dots R_j \dots R_{n-1} R_n \quad (49)$$

And,

$$R_j = \left[ I - \frac{(R_{j+1} R_{j+2} \dots R_n a_j)(R_{n-j+1} \dots R_{n-1} R_n a_j)^T}{(R_{n-j+1} \dots R_{n-1} R_n a_j)^T (R_{n-j+1} \dots R_{n-1} R_n a_j)} \right] \quad (50)$$

Following the same procedure to remove off diagonal entries for the remaining rows of Equation (48), it becomes:

$$\begin{bmatrix} C_1 a_1 & 0 & \ldots & 0 & 0 \\ 0 & C_2 a_2 & \ldots & 0 & 0 \\ \vdots & \vdots & \ddots & \vdots & \vdots \\ 0 & 0 & \ldots & C_{n-1} a_{n-1} & 0 \\ 0 & 0 & \ldots & 0 & C_n a_n \end{bmatrix} x = \begin{bmatrix} C_1 b \\ C_2 b \\ \vdots \\ C_{n-1} b \\ C_n b \end{bmatrix} \qquad (51)$$

With,

$$C_i = R_1 R_2 \ldots R_j \ldots R_{n-1} R_n \exists\, j \neq i \qquad (52)$$

Multiplying both sides of Equation (52) with $A^T{}_n$ to revert back to our least squares solution as in Equation (35),

$$\begin{bmatrix} a^T{}_1 \\ a^T{}_2 \\ \vdots \\ a^T{}_{n-1} \\ a^T{}_n \end{bmatrix} \begin{bmatrix} C_1 a_1 & 0 & \ldots & 0 & 0 \\ 0 & C_2 a_2 & \ldots & 0 & 0 \\ \vdots & \vdots & \ddots & \vdots & \vdots \\ 0 & 0 & \ldots & C_{n-1} a_{n-1} & 0 \\ 0 & 0 & \ldots & 0 & C_n a_n \end{bmatrix} x = \begin{bmatrix} a^T{}_1 \\ a^T{}_2 \\ \vdots \\ a^T{}_{n-1} \\ a^T{}_n \end{bmatrix} \begin{bmatrix} C_1 b \\ C_2 b \\ \vdots \\ C_{n-1} b \\ C_n b \end{bmatrix} \qquad (53)$$

Solving for $x$ in Equation (53), we obtain Equation (38).

$$x = \begin{bmatrix} \frac{a^T{}_1 C_1 b}{a^T{}_1 C_1 a_1} \\ \frac{a^T{}_2 C_2 b}{a^T{}_2 C_2 a_2} \\ \vdots \\ \frac{a^T{}_{n-1} C_{n-1} b}{a^T{}_{n-1} C_{n-1} a_{n-1}} \\ \frac{a^T{}_n C_n b}{a^T{}_n C_n a_n} \end{bmatrix} \qquad \text{(Repeat)}$$

## 6. Proof of the Choice of $C$ Matrix

In this section, we will provide a novel factorization of LS solution which confirms our choice of $C$ matrix. For the sake of brevity, we will provide the proof for $m = 2$. The proof can be similarly extended to any dimension. Substituting $m = 2$ in Equation (26) and expanding,

$$\begin{bmatrix} a^T{}_1 \\ a^T{}_2 \end{bmatrix} [a_1 \quad a_2] x = \begin{bmatrix} a^T{}_1 \\ a^T{}_2 \end{bmatrix} b_2 \qquad (54)$$

$$\begin{bmatrix} a^T{}_1 a_1 & a^T{}_1 a_2 \\ a^T{}_2 a_1 & a^T{}_2 a_2 \end{bmatrix} x = \begin{bmatrix} a^T{}_1 b_2 \\ a^T{}_2 b_2 \end{bmatrix} \qquad (55)$$

Proceeding directly for the solution,

$$x = \frac{1}{(a^T{}_1 a_1)(a^T{}_2 a_2) - (a^T{}_1 a_2)(a^T{}_2 a_1)} \begin{bmatrix} a^T{}_2 a_2 & -a^T{}_1 a_2 \\ -a^T{}_2 a_1 & a^T{}_1 a_1 \end{bmatrix} \begin{bmatrix} a^T{}_1 b_2 \\ a^T{}_2 b_2 \end{bmatrix} \qquad (56)$$

This would yield,

$$x = \frac{1}{(a^T{}_1 a_1)(a^T{}_2 a_2) - (a^T{}_1 a_2)(a^T{}_2 a_1)} \begin{bmatrix} (a^T{}_2 a_2)(a^T{}_1 b_2) - (a^T{}_1 a_2)(a^T{}_2 b_2) \\ (a^T{}_1 a_1)(a^T{}_2 b_2) - (a^T{}_2 a_1)(a^T{}_1 b_2) \end{bmatrix} \qquad (57)$$

$$x = \frac{1}{(a^T{}_1 a_1)(a^T{}_2 a_2) - (a^T{}_1 a_2)(a^T{}_2 a_1)} \begin{bmatrix} \frac{(a^T{}_2 a_2)(a^T{}_1 b_2) - (a^T{}_1 a_2)(a^T{}_2 b_2)}{a^T{}_2 a_2} a^T{}_2 a_2 \\ \frac{(a^T{}_1 a_1)(a^T{}_2 b_2) - (a^T{}_2 a_1)(a^T{}_1 b_2)}{a^T{}_1 a_1} a^T{}_1 a_1 \end{bmatrix} \qquad (58)$$

$$x = \frac{1}{(a^T{}_1 a_1)(a^T{}_2 a_2) - (a^T{}_1 a_2)(a^T{}_2 a_1)} \begin{bmatrix} a^T{}_1 \left[I - \frac{a_2 a^T{}_2}{a^T{}_2 a_2}\right] b_2 (a^T{}_2 a_2) \\ a^T{}_2 \left[I - \frac{a_1 a^T{}_1}{a^T{}_1 a_1}\right] b_2 (a^T{}_1 a_1) \end{bmatrix} \quad (59)$$

The term $(a^T{}_1 a_1)(a^T{}_2 a_2) - (a^T{}_1 a_2)(a^T{}_2 a_1)$ can be in written two ways. Firstly, multiply and divide it by $a^T{}_2 a_2$,

$$(a^T{}_1 a_1)(a^T{}_2 a_2) - (a^T{}_1 a_2)(a^T{}_2 a_1) = \frac{(a^T{}_1 a_1)(a^T{}_2 a_2) - (a^T{}_1 a_2)(a^T{}_2 a_1)}{a^T{}_2 a_2} a^T{}_2 a_2$$
$$(a^T{}_1 a_1)(a^T{}_2 a_2) - (a^T{}_1 a_2)(a^T{}_2 a_1) = a^T{}_1 \left[I - \frac{a_2 a^T{}_2}{a^T{}_2 a_2}\right] a_1 (a^T{}_2 a_2) \quad (60)$$

Secondly, multiply and divide it by $a^T{}_1 a_1$,

$$(a^T{}_1 a_1)(a^T{}_2 a_2) - (a^T{}_1 a_2)(a^T{}_2 a_1) = \frac{(a^T{}_1 a_1)(a^T{}_2 a_2) - (a^T{}_1 a_2)(a^T{}_2 a_1)}{a^T{}_1 a_1} a^T{}_1 a_1$$
$$(a^T{}_1 a_1)(a^T{}_2 a_2) - (a^T{}_1 a_2)(a^T{}_2 a_1) = a^T{}_2 \left[I - \frac{a_1 a^T{}_1}{a^T{}_1 a_1}\right] a_2 (a^T{}_1 a_1) \quad (61)$$

Opting Equation (60) and Equation (61) for first and second rows of Equation (59) respectively,

$$x = \begin{bmatrix} \dfrac{a^T{}_1 \left[I - \frac{a_2 a^T{}_2}{a^T{}_2 a_2}\right] b_2 (a^T{}_2 a_2)}{a^T{}_1 \left[I - \frac{a_2 a^T{}_2}{a^T{}_2 a_2}\right] a_1 (a^T{}_2 a_2)} \\ \dfrac{a^T{}_2 \left[I - \frac{a_1 a^T{}_1}{a^T{}_1 a_1}\right] b_2 (a^T{}_1 a_1)}{a^T{}_2 \left[I - \frac{a_1 a^T{}_1}{a^T{}_1 a_1}\right] a_2 (a^T{}_1 a_1)} \end{bmatrix} \quad (62)$$

Terms $a^T{}_1 a_1$ and $a^T{}_2 a_2$ are scalars. Cancelling and simplifying Equation (62),

$$x = \begin{bmatrix} \dfrac{a^T{}_1 \left[I - \frac{a_2 a^T{}_2}{a^T{}_2 a_2}\right] b_2}{a^T{}_1 \left[I - \frac{a_2 a^T{}_2}{a^T{}_2 a_2}\right] a_1} \\ \dfrac{a^T{}_2 \left[I - \frac{a_1 a^T{}_1}{a^T{}_1 a_1}\right] b_2}{a^T{}_2 \left[I - \frac{a_1 a^T{}_1}{a^T{}_1 a_1}\right] a_2} \end{bmatrix} = \begin{bmatrix} \dfrac{a^T{}_1 C_1 b_2}{a^T{}_1 C_1 a_1} \\ \dfrac{a^T{}_2 C_2 b_2}{a^T{}_2 C_2 a_2} \end{bmatrix} \quad (63)$$

The factorization process reveals the same $C$ matrix we employed in Equation (38) as the solutions obtained by Equation (38) and (63) are exactly same. This confirms our choice of $C$ matrix in Equation (38).

## 7. Algorithm

In this section, we present an algorithm for proposed method. We will start with an example, say computation of $x_1$, and onwards build a generic algorithm. We begin by re-writing Equation (15),

$$R_n A_{n-1} x = R_n b \quad (64)$$

$A_{n-1}$ is generated from $A_n$ by removing its last column. This last column is used to construct $R_n$ which is then multiplied w$A_{n-1}$.

$$R_n A_{n-1} = \left[I - \frac{a_n a_n{}^T}{a_n{}^T a_n}\right] A_{n-1} = \left[A_{n-1} - \frac{a_n}{a_n{}^T a_n} a_n{}^T A_{n-1}\right] = \left[b - \frac{a_n}{a_n{}^T a_n} a_n{}^T b\right] \quad (65)$$

$A_{n-1}$ and $b$ are then updated.

$$A_{n-1} = \left[A_{n-1} - \frac{a_n}{a_n^T a_n} a_n^T A_{n-1}\right] \tag{66}$$

$$b = \left[b - \frac{a_n}{a_n^T a_n} a_n^T b\right] \tag{67}$$

This step can be achieved in a single line of code. It would eliminate the need of extra variables to store their previous values of $A$ and $b$. Similarly, $A_{n-2}$ is generated from $A_{n-1}$ by removing its last column from which $R_{n-1}$ is constructed for subsequent multiplication by $A_{n-2}$ and $b$ and then the update.

$$R_{n-1} A_{n-2} x = R_{n-1} b \tag{68}$$

And,

$$\left[A_{n-2} - \frac{a_{n-1}}{a_{n-1}^T a_{n-1}} a_{n-1}^T A_{n-2}\right] x = b - \frac{a_{n-1}}{a_{n-1}^T a_{n-1}} a_{n-1}^T b \tag{69}$$

$$A_{n-2} = A_{n-2} - \frac{a_{n-1}}{a_{n-1}^T a_{n-1}} a_{n-1}^T A_{n-2} \tag{70}$$

$$b = b - \frac{a_{n-1}}{a_{n-1}^T a_{n-1}} a_{n-1}^T b \tag{71}$$

Continuing in same fashion,

$$a_1 = a_1 - \frac{a_2}{a_2^T a_2} a_2^T a_1 \tag{72}$$

$A_1$ will just be a single column at this step so we have replaced it by $a_1$.

$$b = b - \frac{a_2}{a_2^T a_2} a_2^T b \tag{73}$$

To compute $x_1$, it would be appear from Equation (24) and (25) that there would be a need to store the original $A_n$ matrix.

$$x_1 = \frac{a_1^T b}{a_1^T a_1} \tag{74}$$

But this would be unnecessary. In ordinary least squares,

$$a_1^T e = a_1^T (b - a_1^T x_1) = 0 \tag{75}$$

$a_1$ is orthogonal to $e$ which makes the dot product in Equation (75) equal to zero. In our case, it is the $C_1$ matrix that makes the dot product equal to zero.

$$a_1^T C_1 e = a_1^T C_1 (b - a_1^T x_1) = 0 \tag{76}$$

Hence, multiplication by $a_1$ can be avoided. Here, ordinary summation will do the job.

$$x_1 = \frac{\sum_{i=1}^m b_i}{\sum_{i=1}^m a_i} \tag{77}$$

Equation (77) represents the desired solution. A stepwise algorithm for computing any arbitrary unknown is given in Table 1.

Table 1. Stepwise recursive algorithm for the proposed method

1. Initialize a variable $k$ that represents the $k$-th unkown to be determined.

2. Initialize a variable $p$ such that $p = n$ and ensure $p \neq k$.

3. Generate $A_{p-1}$ from $A_p$ by separating $a_p$ and update the $A_{p-1}$ matrix using:
$$A_{p-1} = \left[ A_{p-1} - \frac{a_p}{a_p^T a_p} a_p^T A_{p-1} \right]$$

4. Update the $b$ vector using:
$$b = \left[ b - \frac{a_p}{a_p^T a_p} a_p^T b \right]$$

5. Decrement $p$.

6. Move back to step 3 and keep iterating until $p = 1$.

7. Compute the unknown using:
$$x_k = \frac{\sum_{i=1}^{m} b_i}{\sum_{i=1}^{m} a_i}$$

## 8. Generation of Inverse Vectors and Comparison with Qr Decomposition

In this section, we will demonstrate that the first run of the algorithm described in section 7 will generate the first inverse vector [27], the first unknown $x_1$, and the complete $Q$ matrix,. We begin by considering Equation (65). In that equation, we removed the last column of $A_n$ to form $A_{n-1}$, constructed $R_n$ from the removed column, multiplied $R_n$ with $A_{n-1}$, and then finally disposed off that last column. We proceeded in the similar fashion until we arrived at $a_1$ in Equation (72). But during the entire process, we disposed off the columns that have been used to construct $R_{n-i}$ matrices at each stage. If these columns are kept instead, then they would constitute the $Q$ matrix and the last column $a_1$ will ultimately be the inverse vector [27]. This can be explained as follows.

Removal of the $n$-th column from $A_n$ will form $A_{n-1}$ as in Equation (65). Removed column is kept as $a_n$. Multiplying $R_n$ with $A_{n-1}$ removes the projections of the columns of $A_{n-1}$ that lie along $a_n$ as described in Equation (65). Last column of the resulting matrix is removed to generate $A_{n-2}$ in Equation (68). This column is kept as $a_{n-1}$. $a_{n-1}$ will be orthogonal to $a_n$. This is because the projection of $a_{n-1}$ that lies along $a_n$ has been removed after multiplication with $R_n$. Continuing in the same fashion, projection of every removed column will be subtracted from the remaining columns of $A_{n-i}$ matrix due to multiplication with $R_{n-(i+1)}$ matrix. In the end, we obtain $a_1$ in Equation (72) from which all its components that lie parallel to separated columns have been removed. As a result, $a_1$ will be orthogonal to all the remaining columns. This is precisely the definition of inverse vector [27]. Hence, $a_1$ is the first inverse vector and the desired solution to the first unknown $x_1$.

Now we will explain the generation of $Q$ and $R$ matrices. From $a_1$, all its projections that lie along the separated $(n-1)$ columns have been removed. Hence, $a_1$ will generate $(n-1)$ zeros when multiplied with the $A_n$ matrix on the left hand side. This will form the first row of a lower triangular matrix with $(n-1)$ zeros. From $a_2$, all the projections that lie along the last $(n-2)$ columns have been removed. This will generate the second row of the lower triangular matrix with $(n-2)$ zeros. Continuing in the same fashion, we arrive at $a_n$ from which no projection has been removed. This will generate the last row of the lower triangular matrix that is full and does not contain any zeros. Given that the removed columns are stored row-wise in a $Q$ matrix, multiplying it with $A_n$ will produce a lower triangular matrix. As we have started from the $n$-the column and the moved backwards to the first column, the matrix generated will be lower triangular. If instead we start from the first column and then move forward to the $n$-th column, the resulting matrix will be upper triangular. This matrix is termed as $R$ matrix in $QR$ decomposition. So $Q^H A$ will be equal to $R$ or $A = QR$. This completes the $QR$ decomposition process.

**9. Computational Complexity**

We begin the analysis of the computational complexity of the algorithm by examining Equation (66) and (67).

$$A_{n-1} = A_{n-1} - \frac{a_n}{a_n^T a_n} a_n^T A_{n-1} \qquad \text{Repeat}$$

$$b = b - \frac{a_n}{a_n^T a_n} a_n^T b \qquad \text{Repeat}$$

The term $a_n^T A_{n-1}$ will require the largest number of multiplications which is $(n-1)m$. In the next step, multiplication by $R_{n-1}$ will require $(n-2)m$ multiplications. Continuing in same fashion to $R_1$, there will be a total of $[1 + \cdots + (n-3) + (n-2) + (n-1)]m$ multiplications. The series $[1 + \cdots + (n-3) + (n-2) + (n-1)]$ can be summed up using Gauss's formula $\frac{n(n+1)}{2}$ for the sum of first $n$ numbers. Since the last number is $(n-1)$, the sum of the series becomes $\frac{n(n-1)}{2}$, resulting in $\frac{n(n-1)}{2} m$ multiplications. So the order of complexity for the computation of a single unknown would be $O\left(\frac{n^3}{2}\right)$ for a large system. It is tempting to think that the computation of all $n$-unknowns would require the order of complexity to rise to $n^4$ but this is not the case. Once an unknown is computed, say $x_1$, it can be substituted backwards which requires only $m$ additional multiplications. Now for the second unknown, the reduced system does not have to be solved all over again as all the $R$'s and $b$'s that are required for its calculation have been computed in the first phase, i.e., the phase in which $x_1$ was calculated. Leaving aside the first and the last unknowns which do not require any back substitution, a total of $(n-2)m$ extra multiplications will be required for computing all unknowns. Therefore, the order of the complexity will not rise above $n^3$. Therefore, the total cost of the algorithm will be $\frac{n^3}{2}$. Comparing it with the QR algorithm which has the cost of $\frac{2n^3}{3}$ [22],

$$\frac{\frac{2n^3}{3} - \frac{n^3}{2}}{\frac{2n^3}{3}} = \frac{\frac{n^3}{6}}{\frac{2n^3}{3}} = \frac{1}{4} = .25$$

There is a 25% reduction in the total computational cost.

**10. Stability**

In this section we analyze the stability of the proposed method. For this purpose, we determine its spectral radius. Spectral radius is defined as the absolute value of the largest eigenvalue of the controlling matrix. It should be less than or equal to one for a stable operation [22]. In the proposed method, matrices that control the solution are rank one-projection matrices. These matrices have a single largest eigenvalue of one whereas all the remaining eigenvalues are zero. Their spectral radius is 1. This means that if an arbitrary vector $x$ is multiplied with these matrices, its length will not be inflated.

$$\|Rx\|^2 = x^T R^T R x = x^T R x = x^T \left[I - \frac{aa^T}{a^T a}\right] x = x^T x - \frac{\|a^T x\|^2}{a^T a} \tag{78}$$

Where,

$$R^T R = I - 2\frac{aa^T}{a^T a} + \frac{aa^T}{a^T a} = I - \frac{aa^T}{a^T a} = R$$

When $a^T x = 0$, Eq. (78) becomes $\|Rx\|^2 = \|x\|^2$. In case $a = x$, $\|Rx\|^2 = 0$. In general:

$$\|Rx\|^2 \leq \|x\|^2 \tag{79}$$

Therefore, there is no risk of explosion in the length of $x$ after repeated multiplications.

## 11. Results/Discussion

We now present the simulation results of the algorithm. The components of the channel matrix are chosen to be independent and identically distributed (IID) Gaussian random variables with a zero mean and unity variance. $m$ is selected equal to $n$ as this refers to multi-user case in M-MIMO systems because both the number of transmitting and receiving antennas become very large. Also when $m = n$, an exact solution is possible and the residue and hence the error in the estimate can be zero. Various matrix sizes have been selected and the results obtained in terms of the norm of residue, norm of the error in estimate and the computational time taken are displayed in Table 2 against the state of the art algorithms available in literature. Codes required for simulation of LSMR and LSQR algorithms have been adopted from the website of Stanford University's System Optimization Laboratory [28] and were used as is. As the proposed method is exact, it achieves zero norms in all the cases. Its columns for residue and error norms are not included in the table. Simulation results demonstrate that LSMR, LSQR, and Kaczmarz algorithms yield much higher norms for both the residue and the error for large matrix sizes. Due to these higher norms, the estimate becomes practically useless despite the fact that LSMR and LSQR are much faster than the proposed algorithm. On the other hand, QR being an exact method achieves zero error/residue norms. But it takes more computation time than the proposed method. Hence, the proposed method is a much better choice in this scenario. It has higher speed as compared to QR method and lower error/residue norms when compared to LSMR, LSQR, and Kaczmarz algorithms.

Table 2. Comparison of the proposed algorithm with the state of the art methods

| $M = N$ | Proposed | QR | LSMR | | | LSQR | | | Kaczmarz 100 iterations | | |
|---|---|---|---|---|---|---|---|---|---|---|---|
| | time | time | $\|r\|$ | $\|e\|$ | Time | $\|r\|$ | $\|e\|$ | time | $\|r\|$ | $\|e\|$ | time |
| $N = 20$ | 0 | 0 | 0.0804 | 0.6418 | 0 | 0.1596 | 1.3585 | 0 | 0.5191 | 1.2066 | 0.0630 |
| $N = 40$ | 0 | 0 | 0.1116 | 1.7462 | 0 | 0.1960 | 1.6275 | 0 | 1.3482 | 1.3463 | 0.1260 |
| $N = 60$ | 0.0156 | 0.0211 | 0.1589 | 2.0725 | 0 | 0.5782 | 1.8263 | 0 | 1.5294 | 2.3033 | 0.2200 |
| $N = 80$ | 0.0469 | 0.0627 | 0.1896 | 2.1318 | 0.0150 | 1.2171 | 2.3114 | 0 | 2.5787 | 3.0479 | 0.4070 |
| $N = 100$ | 0.0781 | 0.1045 | 0.2144 | 2.3678 | 0.0151 | 1.6671 | 3.0958 | 0 | 3.2469 | 4.3231 | 0.5010 |
| $N = 120$ | 0.1094 | 0.1459 | 0.2186 | 2.4746 | 0.0153 | 1.8697 | 3.5757 | 0 | 3.8631 | 5.2422 | 0.7050 |
| $N = 140$ | 0.1563 | 0.2083 | 0.2320 | 2.4962 | 0.0159 | 2.3005 | 3.6616 | 0 | 4.5234 | 9.3310 | 0.9870 |
| $N = 160$ | 0.2188 | 0.2916 | 0.2540 | 2.6058 | 0.0160 | 2.4816 | 3.7103 | 0 | 4.7671 | 11.7761 | 1.2980 |
| $N = 180$ | 0.3281 | 0.4375 | 0.3468 | 2.8010 | 0.0310 | 3.2124 | 3.7700 | 0.0320 | 5.6000 | 14.2675 | 1.6420 |
| $N = 200$ | 0.4375 | 0.5833 | 0.3862 | 2.8263 | 0.0780 | 3.4337 | 3.8408 | 0.1870 | 5.1865 | 23.5984 | 2.0490 |

**Acknowledgements**

Author is grateful to Professor Dianne O'Leary for her valuable comments about the manuscript. Prof. O'Leary is currently serving as a Professor of Computer Science at the Institute for Advanced Computer Studies, University of Maryland at College Park. Her suggestions regarding the reduction in computational complexity of the algorithm were especially invaluable for this manuscript.

**References**
[1] Mietzner J, Schober R, Lampe L, Gerstacker W and Hoeher P. Multiple-antenna techniques for wireless communications - A comprehensive literature survey. *IEEE Communications Surveys and Tutorials.* 2009; 11: 87-105.
[2] Tse D and Viswanath P. Fundamentals of wireless communication. Cambridge University Press Marzetta TL. Noncooperative cellular wireless with unlimited numbers of base station antennas. *IEEE Transactions on Wireless Communications.* 2005; 201(09): 3590-3600.
[3] Rusek F, Persson D, Buon Kiong L, Larsson EG, Marzetta TL, et al. Scaling Up MIMO: Opportunities and Challenges with Very Large Arrays. *IEEE Signal Processing Magazine.* 2013; 30: 40-60.
[4] Vishwanath S, Jindal N, Goldsmith A. Duality, achievable rates, and sum-rate capacity of Gaussian MIMO broadcast channels. *IEEE Transactions on Information Theory.* 2003; 49: 2658-2668.
[5] Weingarten H, Steinberg Y, Shamai S. The capacity region of the Gaussian multiple-input multiple-output broadcast channel. *IEEE Transactions on Information Theory.* 2006; 52: 3936-3964.
[6] Larsson EG. MIMO detection methods: how they work. *IEEE Signal Processing Magazine.* 2009; 26: 91-95.
[7] Jalden J, Ottersten B. On the complexity of sphere decoding in digital communications. *IEEE Transactions on Signal Processing.* 2005; 53: 1474-1484.
[8] Peel CB, Hochwald BM, Swindlehurst AL. A vector-perturbation technique for near-capacity multiantenna multiuser communication-part I: channel inversion and regularization. *IEEE Transactions on Communications.* 2005; 53: 195-202.
[9] Ryan DJ, Collings IB, Clarkson IVL, Heath RW Jr. Performance of vector perturbation multiuser MIMO systems with limited feedback. *IEEE Transactions on Communications.* 2009; 57: 2633-2644.
[10] Windpassinger C, Fischer RFH, Huber JB. Lattice-reduction-aided broadcast precoding. *ITG-Fachbericht.* 2003; 403-408.
[11] Liang YC, Cheu EY, Bai L, Pan G. On the relationship between MMSE-SIC and BI-GDFE receivers for large multiple-input multiple-output channels. *IEEE Transactions on Signal Processing.* 2008; 56: 3627-3637.
[12] Couillet R, Debbah M. Random Matrix Methods for Wireless Communications. Cambridge University Press. 2011.
[13] Edman F, Owall V. *Implementation of a scalable matrix inversion architecture for triangular matrices.* 14th IEEE International Symposium on Personal, Indoor and Mobile Radio Communications Proceedings. Piscataway, NJ, USA: IEEE. 2003; 2558-2562.
[14] Golub G, Van Loan C. Matrix Computations (Johns Hopkins Studies in Mathematical Sciences) (3rd Edition). The Johns Hopkins University Press. 1996.
[15] Honig ML, Weimin X. Performance of reduced-rank linear interference suppression. I*EEE Transactions on Information Theory.* 2001; 47: 1928-1946.
[16] Karkooti M, Cavallaro JR, Dick C. *FPGA implementation of matrix inversion using QRD-RLS algorithm.* 39th Asilomar Conference on Signals, Systems and Computer. Piscataway, NJ, USA: IEEE. 2005; 1625-1629.
[17] Ma L, Dickson K, McAllister J, McCanny J. QR decomposition-based matrix inversion for high performance embedded MIMO receivers. *IEEE Transactions on Signal Processing.* 2011; 59: 1858-1867.
[18] Muller RR, Verdu S. Design and analysis of low-complexity interference mitigation on vector channels. *IEEE Journal on Selected Areas in Communications.* 2001; 19: 1429-1441.
[19] Stewart GW (1998) Matrix Algorithms: Volume 1: Basic Decompositions. SIAM.
[20] Wu D, Eilert J, Liu D, Wang D, Al-Dhahir N, et al. (2007) Fast complex valued matrix inversion for multi-user STBC-MIMO decoding. *IEEE Computer Society Annual Symposium on VLSI: Emerging VLSI Technologies and Architectures, ISVLSI'07*, March 9, 2007 - March 11, 2007. Porto Alegre, Brazil: Inst. of Elec. and Elec. Eng. Computer Society. pp. 325-330.
[21] Strang G. Computational Science and Engineering. Wellesley-Cambridge Press. 2007.
[22] Fong D and Saunders M. LSMR: An iterative algorithm for sparse least-squares problems. *SIAM Juornal on Scientific Computing.* 2011; 2950-2971.
[23] Paige CC and Saunders MA. LSQR: An Algorithm for Sparse Linear Equations and Sparse Least Squares. *ACM Trans Math Softw.* 1982; 8: 43-71.
[24] Strohmer T, Vershynin R. A Randomized Kaczmarz Algorithm with Exponential Convergence. *J Fourier Anal Appl.* 2009; 15: 262-278.
[25] Fleisch D. A Student's Guide to Vectors and Tensors. Cambridge University Press. 2011.
[26] Anjum MAR, Ahmed MM. *A New Approach for Inversion of Large Random Matrices in Massive MIMO Systems.* PLoS ONE. 2014.
[27] http://www.stanford.edu/group/SOL/software/lsmr.html.